\documentclass[prd,a4paper,superscriptaddress,nofootinbib,10pt]{revtex4} 
\usepackage{graphicx}
\usepackage{amssymb}
\usepackage{amsmath}
\usepackage{lscape}
\usepackage{mathrsfs}
\usepackage{textcomp}
\usepackage{epsfig}
\usepackage{slashed}
\usepackage{comment}
\usepackage{xcolor}
\usepackage{ulem}
\renewcommand{\d}{\mathrm{d}}

\begin{document}
\begin{flushright}
ZTF-EP-19-02
\end{flushright}
\title{Noncommutativity and the Weak Cosmic Censorship}

\author{Kumar S. Gupta}
\email{kumars.gupta@saha.ac.in}
\affiliation{Theory Division, Saha Institute of Nuclear Physics, 1/AF Bidhannagar, Calcutta 700064, India}

\author{Tajron Juri\'c}
\email{tjuric@irb.hr}

\author{Andjelo Samsarov}
\email{asamsarov@irb.hr}
\affiliation{Rudjer Bo\v{s}kovi\'c Institute, Bijeni\v cka  c.54, HR-10002 Zagreb, Croatia}

\author{Ivica Smoli\'c}
\email{ismolic@phy.hr}
\affiliation{Department of Physics, Faculty of Science, University of Zagreb, 10000 Zagreb, Croatia}

\date{\today}

\begin{abstract}

We show that a  noncommutative massless scalar probe can dress a naked singularity in $AdS_3$ spacetime, consistent with the weak cosmic censorship. The dressing occurs at high energies, which is typical at the Planck scale. Using a noncommutative duality, we show that the dressed singularity has the geometry of a rotating BTZ black hole which satisfies all the laws of black hole thermodynamics. We calculate the entropy and the quasi-normal modes of the dressed singularity and show that the corresponding spacetime can be quantum mechanically complete. The noncommutative duality also gives rise to a light scalar, which can be relevant for early universe cosmology. 

\end{abstract}

\maketitle

%%%%%%%%%%%%%%%%%%%%%%%%%
\section{Introduction}%%%
%%%%%%%%%%%%%%%%%%%%%%%%%

Gravitational singularities are mainly considered in the context of general relativity, where density  becomes infinite at the center of a black hole, and within astrophysics and cosmology as the earliest state of the universe during the Big Bang \cite{hawkingpenrose}. 
It is expected that any object undergoing a gravitational collapse, compressed beyond its  Schwarzschild radius would form a black hole, inside which a singularity  would be formed, which is covered by an event horizon. This is in accordance with the weak cosmic censorship hypothesis \cite{Penrose:1969pc,Penrose:1900mp,Penrose:1999vj, hawking}, which states that naked singularities should not appear in nature. Otherwise the causality may break down and physics, in particular general relativity, may lose its predictive power.\\

 There exists a strong belief that quantum theory could help with smearing out of classical singularities. This has been discussed in the context of loop quantum gravity \cite{Ashtekar:2003hd,Ashtekar:2005qt,Bojowald:2006rp,Yonika:2017qgo,Saini:2016vgo}, string theory \cite{sumit}, fuzzballs \cite{mathur}, higher derivative gravity \cite{anupam} and asymptotic safety \cite{asympt}. It has also been argued that the general framework of quantum mechanics helps to address the singularity problem \cite{,Saini:2014qpa,Greenwood:2008ht,Wang:2009ay}. In particular, there are examples of static spacetimes with timelike singularities where the dynamics of a  quantum test particle is completely well behaved for all time \cite{horowitz}.  Another problem that is closely related to the existence of singularities is that of geodesic completeness \cite{hofmann}.
In this context it is worthy to mention that  although the spacetime may be geodesically incomplete, 
 the  Laplacian can still be essentially self-adjoint, giving rise to a unique evolution in time \cite{horowitz, hofmann, reedsimon, chernof, isibasi, isibasi1, konkowski, wald, Konkowski:2016pgi}.\\

 The 2+1 dimensional spacetime with a negative cosmological constant and the associated BTZ geometry \cite{banados,heno} provides a convenient setting to investigate various concepts related to singularity resolution. Here one may  consider quantum backreaction effects as a way to put singularity under control  \cite{Martinez:1996uv,recent,recent1}. Indeed, it has recently been shown that quantum backreaction can dress a naked singularity in 2+1 dimensions with a negative cosmological constant \cite{recent,recent1}. In another approach, it was found  that the BTZ black hole cannot be spun-up beyond a certain  limit, confirming the weak cosmic censorship in 2+1 dimensional asymptotically AdS spacetime \cite{Rocha:2011wp}. Since a class of counterexamples  have been identified  in Einstein-Maxwell gravity that violates the cosmic censorship hypothesis, an idea was put forth that a weak gravity conjecture \cite{weakgravity}  might fix the issue and
preserve cosmic censorship \cite{Crisford:2017gsb,Horowitz:2019eum}.\\

 In this paper we explore the role of spacetime noncommutativity within the context of the weak cosmic censorship hypothesis. It is well known that noncommutativity is a possible description of the spacetime structure at the Planck scale, which arises naturally when general relativity and the quantum uncertainty principle are considered in unison \cite{ahluwalia,dop1,dop2}. The idea of resolution of the spacetime singularities have already been discussed in fuzzy manifolds \cite{buric1,buric2}. Here we take the $\kappa$-deformed algebra as a model of spacetime noncommutativity \cite{kappa1,kappa2,kappa3}, which arises in several descriptions of noncommutative black holes \cite{schupp,al1,al2} and cosmology \cite{ohl}.  For this purpose we employ a toy model consisting of a naked singularity in 2+1 gravity \cite{banados,heno,pitelli}. We will show that probing such a spacetime with a massless noncommutative scalar field transforms the naked singularity into BTZ black hole. This happens when the frequency or the energy of the NC probe satisfies a suitable condition described by the system parameters. The features of the black hole depends on the noncommutative parameter as well as the energy scale of the probe.  In addition, this process produces a very light massive scalar field whose mass depends on the noncommutative deformation parameter. Noncommutative geometry has been used to describe cosmological models of early universe \cite{matilde} including noncommutative inflation \cite{brand,mota}. It is plausible that the light scalar predicted by our analysis can be of relevance in early universe cosmology within the context of quintessence \cite{quint}. It may be noted that the question of quantum completeness of the BTZ spacetime using a noncommutative probe was considered in \cite{Juric:2018qzl}, which suggested that noncommutativity  may smear out the singularity by allowing quantum completeness for a wider range of BTZ parameters.\\

The paper is organized as follows. In section 2 we summarize the basic arguments leading to a dual picture between a system including noncommutative scalar field that probes a classical nonrotating BTZ geometry and a system consisting of a spinning BTZ black hole perturbed by a massive commutative scalar field \cite{ncbtz,ncbtz1,ncbtz2,ncbtz3}. These two systems appear to be mathematically equivalent, with the equivalence being enforced by the parameter transformations expanded up to first order in deformation. In section 3,  we analyze conditions under which noncommutativity is able to dress the naked singularity. For this we start with a naked singularity in 2+1 dimensional AdS spacetime and probe it with a noncommutative massless scalar field. We show that for a suitable range of the probe frequency and the system parameters, the naked singularity is dressed as a BTZ black hole. In Section 4 we study the physical properties of the dressed singularity and conclude the paper in Section 5 with some final remarks.\\

%%%%%%%%%%%%%%%%%%%%%%%%%%%%%%%%%%%%%%%%%%%%%%%%%%%%%%%%%%%%%%
\section{BTZ black hole and noncommutativity}%%%
%%%%%%%%%%%%%%%%%%%%%%%%%%%%%%%%%%%%%%%%%%%%%%%%%%%%%%%%%%%%%%
\subsection{Noncommutative framework}

Before describing a model on which our analysis is based, we  recall  the formal setting utilized for building the model.
Hence, starting with some noncommutative (NC) algebra of coordinates $\hat{x}_{\mu},$
we may consider its universal enveloping algebra, denoted by $\cal \hat{A}$, which consists of  formal power series  expansions  in coordinates  $\hat{x}_{\mu}.$ More precisely, introducing an algebra $\cal{A}$ with  unit element $1,$ and generated by commutative coordinates $x_{\mu}$    and  yet  introducing the action 
$\triangleright: \mathcal{H}\mapsto\mathcal{A},$ one may identify the algebra $\cal{A}$ as a $D$-module over the ring of polynomials in $\hat{x}_{\mu}.$
 The action 
$\triangleright: \mathcal{H}\mapsto\mathcal{A}$ itself  is defined by
\begin{equation}\label{djelovanje}
x_{\mu} \triangleright 1=x_{\mu},\quad p_{\mu}\triangleright 1=0 \quad   \mbox{and} \quad x_{\mu} \triangleright f(x)=x_{\mu}f(x),\quad p_{\mu}\triangleright f(x)=i\frac{\partial f}{\partial x^{\mu}}.
\end{equation} 
Here, $x_{\mu}$ and $p_{\mu}$ are the generators of the Heisenberg algebra $\mathcal{H}$  satisfying the relations,
\begin{equation}\label{H}
[x_{\mu},x_{\nu}]=[p_{\mu},p_{\nu}]=0, \quad [p_{\mu},x_{\nu}]=i\eta_{\mu\nu},
\end{equation}
where $\eta_{\mu\nu}=\text{diag}(-,+,+)$.
Moreover, there exists an isomorphism between the NC algebra $\cal \hat{A}$,
generated by the noncommutative coordinates 
$\hat{x}_{\mu}$ and the commutative algebra $\cal A_{\star}$,
generated by the commutative coordinates  
$x_{\mu}$, but this time with $\star$ as the algebra multiplication. The star product  between any two elements $f(x)$ and $g(x)$ 
in $\cal A_{\star}$ is defined as
\begin{equation}\label{star}
f(x)\star g(x)=\hat{f}(\hat{x})\hat{g}(\hat{x})\triangleright 1,
\end{equation}
where $\hat{f}(\hat{x})$ and $\hat{g}(\hat{x})$ are the elements in $\cal \hat{A}$
that are respectively and uniquely assigned to the elements $f(x)$ and $g(x),$
 through the following correspondences, 
$\hat{f} (\hat{x}) \triangleright 1 = f(x), \quad \text{and} \quad \hat{g} (\hat{x})
\triangleright 1 = g(x). \; $  
These correspondences provide a basis for the isomorphism between  $\cal \hat{A}$ and   $\cal A_{\star}$.\\

Furthermore, we let the coordinates $\hat{x}_\mu $ satisfy the $\kappa$-Minkowski algebra \cite{kappa1,kappa2,kappa3}
\begin{equation} \label{kappaminkowski}
  [\hat{x}_{\mu}, \hat{x}_{\nu}] = i(a_{\mu} \hat{x}_{\nu}  - a_{\nu} \hat{x}_{\mu} ).
\end{equation}
Here $a_{\mu} $  are the components of a deformation 3-vector in Minkowski space with the 
%signature 
metric  $\eta_{\mu\nu}$. 
%$\eta_{\mu\nu}=\text{diag}(-,+,+)$.
 We shall be concerned with the situation where the deformation vector is oriented in a time direction, $a_{\mu} = (a,0,0),$ where  $a$  is the deformation parameter, $a=\frac{1}{\kappa},$ that
 fixes the energy
scale at which the effects of noncommutativity are expected  to occur. Most frequently it is
taken to be of the order of the Planck length.\\

It is important to emphasize that the  coordinates  $\hat{x}_{\mu}$
admit a differential operator representation within the enveloping algebra of
$ \mathcal{H} $  in terms of the formal power series in $x_\mu$ and $p_\nu$ \cite{Meljanac:2007xb, Govindarajan:2008qa, Juric:2013foa, Juric:2015aza}.\\

%%%%%%%%%%%%%%%%%%%%%%%%%%%%%%%%%%%%%%%%%%%%%%%%%%%%%%%%%%%%%%%%%%%%%%%%%%%%%%%%%%%%%%%%%%%%%%%%%%%%%%%
\subsection{NC scalar field in BTZ background and mapping to equivalent commutative model}%%%
%%%%%%%%%%%%%%%%%%%%%%%%%%%%%%%%%%%%%%%%%%%%%%%%%%%%%%%%%%%%%%%%%%%%%%%%%%%%%%%%%%%%%%%%%%%%%%%%%%%%%%%

In view of the setting just described, in \cite{ncbtz, ncbtz1, ncbtz2, ncbtz3} a model was proposed that is based on the action
\begin{equation}\begin{split} \label{action}
\hat{\mathcal{S}}&=\int \text{d}^{3}x\sqrt{-g}\ \ g^{\mu\nu}\left(\partial_{\mu}\phi\star\partial_{\nu}\phi\right)\\
&=\int \text{d}^{3}x\sqrt{-g}\ \ g^{\mu\nu}\left(\partial_{\mu}\hat{\phi}\partial_{\nu}\hat{\phi}\triangleright 1\right)\\
\end{split}\end{equation}
 describing  a
coupling of the scalar particle to a metric of the form (in units
where $8\pi G =c=\hbar=1$)
\begin{equation}\label{btzmetric}
g_{\mu\nu}=\begin{pmatrix}
M - \frac{r^2}{l^2} &0&0\\
0&\frac{1}{\frac{r^2}{l^2}-M}&0\\
0&0&r^2\\
\end{pmatrix},
\end{equation}
where $l$ is related to the cosmological constant $\Lambda$ as $l =
  \sqrt{-\frac{1}{\Lambda}}$. 
As the noncommutative coordinates used in constructing the action (\ref{action}) satisfy  $\kappa$-Minkowski algebra, it is understood that this action is invariant under  $\kappa$-deformed Poincar\'{e} symmetry, at least at the semi-classical level.
This means that the gravity  is considered to be described by the classical degrees of freedom and is not affected by the spacetime deformation, $(\delta_{a} g_{\mu \nu} =0) $. The matter field instead is subject to a quantum deformation and is consequently described by  a noncommutative degree of freedom.
Thus, the only way  the noncommutativity enters the formalism is through a scalar
field, which  is treated as a noncommutative variable.
This approach therefore amounts to considering a noncommutative  scalar field coupled to a classical geometrical background produced by the spinless $(J=0)$ BTZ black hole with mass $M$.\\

The presence of the  $\kappa$-deformed Poincar\'{e} symmetry in the action (\ref{action}) in a way described above
is ensured by means of implementing the following star product 
\begin{equation} \label{starproduct}
f(x)\star g(x)=f(x)g(x)+i\beta' (\eta^{\mu \nu} x_{\mu} \frac{\partial f}{\partial x^{\nu}})(\eta^{\lambda \sigma} a_\lambda  \frac{\partial g}{\partial x^{\sigma} })+i\beta(\eta^{\mu \nu} a_\mu x_\nu)(\eta^{\lambda \sigma} \frac{\partial f}{\partial x^{\lambda}} \frac{\partial g}{\partial x^{\sigma}})+i\bar{\beta} (\eta^{\mu \nu} a_\mu  \frac{\partial f}{\partial x^{\nu}})(\eta^{\lambda \sigma} x_\lambda  \frac{\partial g}{\partial x^{\sigma}}).
\end{equation}
which is compatible with   $\kappa$-deformed Poincar\'{e} symmetry. 
Here $\beta', \beta$ and $\bar{\beta}$ are the parameters determining the
 differential operator representation of the $\kappa$-Minkowski  algebra and the star product has been expanded up to first order in $a$.
% \cite{kappa}.
It should be noted that  each  choice of  the triplet of parameters $\beta', \beta$ and  $\bar{\beta}$ corresponds to a specific differential operator representation \cite{Meljanac:2007xb}  of noncommutative coordinates and also to a
different choice of the coproduct, as well as  different basis of $\kappa$-Poincar\'{e} algebra. All these notions in turn correspond to the vacuum of the theory and this  should, at least in principle, be fixed by experiment.\\

As shown in \cite{ncbtz}, the leading terms in the action  (\ref{action}) that are consistent with the approximation used and are within a first order in the deformation parameter $a$ do not acquire the contributions from the $\beta'$ and $\bar{\beta}$ terms in the star product (\ref{starproduct}). Thus the only contributions to the action  within a given approximation  come from the $\beta$ term and they give rise to a field equation (Klein-Gordon equation) for a scalar field that   may be written in a generic form
\begin{equation} \label{1}
  ({\Box_g} + {\mathcal{O}}(a)) \phi = 0,
\end{equation}
where $\Box_g$ is a standard Klein-Gordon operator for the metric (\ref{btzmetric}) and ${\mathcal{O}}(a)$ groups together all the remaining terms generated by the star product (\ref{starproduct}) and allowed by the approximation used.
After a separation of the variables, $\phi(r,\theta, t)=R(r)e^{-i\omega t}e^{im\theta}$,the equation (\ref{1}) leads  to the radial equation \cite{ncbtz, ncbtz1}
\begin{equation} \label{eomradial}
r\left(M-\frac{r^2}{l^2}\right)\frac{\partial^2 R}{\partial r^2}+\left(M-\frac{3r^2}{l^2}\right)
\frac{\partial R}{\partial r}+\left(\frac{m^2}{r}-\omega^2\frac{r}{\frac{r^2}{l^2}-M}-a\beta\omega\frac{8r}{l^2}\frac{\frac{3r^2}{2l^2}-M}{\frac{r^2}{l^2}-M}\right)R=0,
\end{equation}
where  $\omega$ and $m$ are respectively the energy and the angular momentum
(magnetic quantum number) of the scalar particle.
As already explained,  $\beta $ is the parameter determining the differential operator representation of the $\kappa$-Minkowski algebra. 
Further details are elaborated in
\cite{ncbtz, ncbtz1, ncbtz2, ncbtz3}.\\

Using the substitution
\begin{equation} \label{2}
z=1-\frac{Ml^2}{r^2},
\end{equation}
Eq.(\ref{eomradial}) can be brought  \cite{ncbtz, ncbtz1} into the form
\begin{equation}\label{eom}
z(1-z)\frac{\d^2 R}{\d z^2}+ (1-z)\frac{\d R}{\d z} + \left(\frac{A}{z}+B+\frac{C}{1-z} \right)R=0,
\end{equation}
where the constants $A,B$ and $C$ are
\begin{equation} \label{coefs}
A=\frac{\omega^2 l^2}{4M}+a\beta\omega, \quad B=-\frac{m^2}{4M}, \quad C=3a\beta\omega.
\end{equation}
The equation \eqref{eom}, together with the coefficients \eqref{coefs}, describes the dynamics of a massless NC scalar field with energy $\omega$ and angular momentum $m$, probing a geometry of the BTZ black hole with mass $M$ and vanishing angular momentum ($J=0$). However, as noted in \cite{ncbtz1, ncbtz2}, the form of this equation is the same as the form of the radial equation governing a commutative massive scalar field in a background of a BTZ black hole with a modified mass and nonvanishing angular momentum \cite{Birmingham:2001hc}. Formally, there is another coordinate transformation, given by $z \mapsto w = 1/z$, which preserves the form of the equation (\ref{eom}). However, the latter transformation simultaneously shifts the coordinate position of the black hole horizon from $z = 1$ to $w = \infty$, so that there is no ambiguity left in identification of the parameters $A$, $B$ and $C$. The modified black hole parameters obtained in this way we label respectively by $M'$ and $J'$ and the equivalent black hole setting they describe, we refer to as the dual black hole setting.  The dual black hole parameters  %with respect to the original setting where the NC scalar field probes the background of the BTZ black hole with vanishing angular momentum 
were calculated in \cite{ncbtz2}. Specifically, within a first order of deformation $a,$ 
the mass $M'$ and the angular momentum $J'$ in the dual black hole setting  are given as
\begin{eqnarray} \label{mprimejprime}
  M' & =& M \bigg[ 1 - 4L_{NC} \omega M \bigg( \frac{1}{\lambda} -\frac{2}{l^2 \omega^2 - m^2}
    +\frac{l^2}{\lambda^{2}} \frac{2\sigma^{2} \lambda^{2} -\sigma^{2} l^{2} + \lambda^{2} }{\sigma^{2} l^{2} + \lambda^{2}} \bigg( \frac{1}{l^2 \omega^2 - m^2}
      - \frac{1}{\lambda} \bigg) \bigg) \bigg], \nonumber \\
  J' &=& 4 L_{NC} \omega M^{2} \frac{l^{2}}{\lambda \sigma}   
      \frac{2\sigma^{2} \lambda^{2} -\sigma^{2} l^{2} + \lambda^{2} }{\sigma^{2} l^{2} + \lambda^{2}} 
   \bigg(\frac{1}{\lambda}  -\frac{1}{l^2 \omega^2 - m^2}
          \bigg).
\end{eqnarray}
The quantities $\lambda$, $\sigma$ and $L_{NC}$ appearing in the above two relations are the abbreviations  defined as follows
\begin{eqnarray} \label{5}
 \lambda & \equiv & l^2 \omega^2 + m^2, \\
  \sigma & \equiv & 2 \omega m,\\
	L_{NC} & \equiv & -a\beta>0.
\end{eqnarray}
 One may easily see that when the noncommutative parameter $a$ goes to zero, two situations coincide, i.e. $M'$ goes to $M$  and $J'$ vanishes, as expected. 
 Another interesting  point   is that in the dual black hole setting the scalar field  acquires  a mass $\mu',$
\begin{equation} \label{8}
3L_{NC}\omega=\frac{\mu^{\prime 2}l^2}{4}.  
\end{equation}
Moreover it brings about certain 
  back-reaction effects by affecting the geometry through which it propagates, as being manifestly seen 
by the change in both the mass and the spin of the BTZ black hole.\\

 % In accordance with the above observations,  may
These features may give a spacetime noncommutativity a novel perspective in which noncommutativity acquires two different roles, one that it might be responsible for generating
a particle mass and the other that it may  provide a suitable ground for some back-reaction effects to take place. Moreover, the scalar field that we obtained is very light and, as will be more obvious in the next section, it is a byproduct
 of the dressing via duality. We note that such a very light scalar field may be relevant for quintessence or for
 fuzzy dark matter scenario in early universe cosmology  \cite{quint}.\\

The above analysis implies that the equation (\ref{1}) can be recast into the form
\begin{equation} \label{7}
  \left({\Box_{g'}} - \mu^{\prime 2} \right)\phi = 0,
\end{equation}
 describing a dynamics of the massive commutative scalar field in the metric
\begin{equation}\label{eqbtzmetric}
g'_{\mu\nu}=\begin{pmatrix}
 M' - \frac{r^2}{l^2}- \frac{{(J')}^2}{4 r^2} &0& \frac{-J'}{2}\\
 0&\frac{1}{\frac{r^2}{l^2} + \frac{J'^2}{4 r^2} -M'}&0\\
\frac{-J'}{2} &0& r^2\\
\end{pmatrix}.
\end{equation}
The metric (\ref{eqbtzmetric}) corresponds to a spinning BTZ black hole with the mass $M'$ and the angular momentum $J'$
and $\Box_{g'}$ is the Klein-Gordon operator for this metric.
 Therefore, we see that there exists a mathematical mapping between two physically different situations,
     one describing the  NC massless scalar field in the non-rotational BTZ background, and the other describing
      the ordinary massive scalar field which  probes a BTZ geometry with a non-vanishing angular momentum. 
For further details regarding the interpretation of this analytic mapping and a corresponding equivalence/duality between two physical settings we refer the reader to reference \cite{ncbtz2}.\\

      As expected, the mass and the angular momentum of  the dual spinning BTZ black hole
 are related to its outer and   inner radii,  $\; r'_+$ and $ r'_- ,\;$ through
\begin{equation}
M'=\frac{r'^{2}_{+}+r'^{2}_{-}}{l^2}, \quad  |J'| = \frac{2r'_{+}r'_{-}}{l}.
\end{equation}
For the original non-rotational BTZ  the actual radii are
respectively given by $r_+ = l\sqrt{M}$ and  $r_- =0 $.\\

%%%%%%%%%%%%%%%%%%%%%%%%%%%%%%%%%%%%%%%%%%%%%%
\section{Dressing the naked singularity}%%%
%%%%%%%%%%%%%%%%%%%%%%%%%%%%%%%%%%%%%%%%%%%%%%

As discussed before, there is an equivalence between the system of NC scalar field probing a commutative BTZ background ($M$, $J=0$)  and a commutative scalar field probing an effective BTZ background ($M'=M'(M,a)$, $J'=J'(M,a)$). This effective background captures the NC effect. In this section we consider the case where we start with a naked sigularity ($M<0$). Using the above mentioned duality, we will show that this naked singularity will be dressed and protected by the NC corrections under certain conditions.\\

We start with the assumption that
\begin{equation}\label{aaa}
M<0, \quad M=-\left|M\right|
\end{equation}
and rewrite the condition in \eqref{mprimejprime} for the parameters of the dual effective metric \eqref{eqbtzmetric} as
\begin{equation}\label{rewrite}
M'=M(1+Mf), \quad J'=gM^2
\end{equation}
where
\begin{equation}
f=-4L_{NC} \omega \bigg( \frac{1}{\lambda} -\frac{2}{l^2 \omega^2 - m^2}
    +\frac{l^2}{\lambda^{2}} \frac{2\sigma^{2} \lambda^{2} -\sigma^{2} l^{2} + \lambda^{2} }{\sigma^{2} l^{2} + \lambda^{2}} \bigg( \frac{1}{l^2 \omega^2 - m^2}
      - \frac{1}{\lambda} \bigg) \bigg) \ ,
\end{equation}
\begin{equation} \label{eqng}
g=4 L_{NC} \omega \, \frac{l^{2}}{\lambda \sigma}   
      \frac{2\sigma^{2} \lambda^{2} -\sigma^{2} l^{2} + \lambda^{2} }{\sigma^{2} l^{2} + \lambda^{2}} 
   \bigg(\frac{1}{\lambda}- \frac{1}{l^2 \omega^2 - m^2} \bigg).
\end{equation}
Now we see that even though the initial mass may be negative ($M<0$), the singularities in the effective metric \eqref{eqbtzmetric} will be surrounded by horizon(s) if the conditions
\begin{equation}
M' > 0 \qquad \textrm{and} \qquad |J'| \le M' l
\end{equation}
are fulfilled. These in turn imply the constraints given by
\begin{equation}\label{uvjet}
f>\frac{1}{\left|M\right|} \qquad \textrm{and} \qquad f \ge \frac{|g|}{l} + \frac{1}{|M|} \ .
\end{equation}
Note that if $g\ne 0$ the first constraint is automatically included within the second one, and thus does not present any additional requirement.\\

In order to analyze these conditions further, we write $f$ as
\begin{equation}
\label{eqnf}
f = -4L_{NC} \, \omega \left( \frac{1}{l^2 \omega^2 + m^2} - \frac{2}{l^2 \omega^2 - m^2} \right) + \frac{\sigma}{\lambda}\,g \ ,
\end{equation}
where the function $g$ can be written as
\begin{equation}
g = -4\,\frac{L_{NC} \omega l^2}{\lambda\sigma} \, \frac{P}{\sigma^2 l^2 + \lambda^2} \left( \frac{1}{l^2 \omega^2 - m^2} - \frac{1}{l^2 \omega^2 + m^2} \right)
\end{equation}
with 
\begin{equation}
P = (8m^2\omega^2 + 1) \omega^4 \, l^4 + 2(8 m^2 \omega^2 - 1) m^2 \omega^2 \, l^2 + (8m^2\omega^2 + 1)m^4. \ 
\end{equation}
If we take $P$ as a quadratic polynomial in $l^2$, and note that the associated discriminant $-2(2m\omega)^6$ is nonpositive, while the leading coefficient $(8m^2\omega^2 + 1) \omega^4$ is nonnegative, it follows that $P \ge 0$. In other words, large fraction that constitutes the central part of the function $g$ is nonnegative, which significantly simplifies the analysis. This allows us to write function $g$ as
\begin{equation}
g = -2\,\frac{L_{NC} l^2}{m} \, \frac{\big( 8m^2\omega^2 (l^2\omega^2 + m^2) + l^2\omega^2 - m^2 \big)^2 + 32 m^6 \omega^2}{(l^2 \omega^2 + m^2)(8m^2 \omega^2 + 1)(4l^2 m^2 \omega^2 + (m^2 + l^2 \omega^2)^2)} \left( \frac{1}{l^2 \omega^2 - m^2} - \frac{1}{l^2 \omega^2 + m^2} \right) \ .
\end{equation}
Here we have following subcases:

\begin{itemize}
\item[(a)] $g > 0$ holds if either $m > 0$ and $\omega < m/l$ or $m < 0$ and $\omega > |m|/l$;

\item[(b)] $g < 0$ holds if either $m > 0$ and $\omega > m/l$ or $m < 0$ and $\omega < |m|/l$.
\end{itemize}

Now, suppose that $0 < \omega < |m|/l$. By the analysis above, this implies that $mg > 0$. Furthermore, the first term of $f$ in \eqref{eqnf} is given by
\begin{equation}
-4L_{NC} \, \omega \, \frac{-(l^2 \omega^2 + 3m^2)}{(l^2 \omega^2 + m^2)(l^2 \omega^2 - m^2)},
\end{equation}
and is thus negative. Also, as $(l\omega \pm m)^2 \ge 0$, it follows that
\begin{equation}
\frac{2\omega |m|l}{l^2\omega^2 + m^2} \le 1 \ .
\end{equation}
In conclusion, $0 < \omega < |m|/l$ implies that $f < |g|/l$, so that in this case inequalities \eqref{uvjet} cannot be fulfilled and the dressing of singularity does not occur.\\

The opposite case, when $\omega > |m|/l$, is pretty nontrivial to treat exactly, and one way to extract simple conclusion is to look for the expansions in the limiting cases of the energy $\omega$. For example, if we look at the cases when $\alpha = l\omega/|m| \gg 1$, using expansions
\begin{equation}
f = \frac{4L_{NC}}{l|m|} \left( \frac{1}{\alpha} + O(1/\alpha^3) \right) \ , \quad f - \frac{g}{l} = \frac{4L_{NC}}{l|m|} \left( \frac{1}{\alpha} + O(1/\alpha^3) \right),
\end{equation}
we may deduce that in the following region of probe energies,
\begin{equation}\label{ec}
\frac{|m|}{l} \ll \omega < \frac{4L_{NC}|M|}{l^2}
\end{equation}
the noncommutative dressing is effective.\\

The above approximation is physical and well justified. Namely,  at the Planck scale, the physical processes occur at very high energies. Thus, both the frequency and the angular momentum of the NC probe would typically be very high. Let us consider the probes satisfying the condition
\begin{equation} \label{lowerlim}
l^2 \omega^2 \gg m^2,
\end{equation}
such that
\begin{equation}
 f = 4 \frac{L_{NC}}{l^2 \omega} \quad \textrm{and} \quad g=0.
\end{equation}
 The inequality \eqref{uvjet} on $f$ gives
\begin{equation} \label{Mcond}
  \frac{l^2 \omega}{4 L_{NC}} < |M|.
\end{equation}
Using the eqns. \eqref{lowerlim} and \eqref{Mcond} we find again the inequality  \eqref{ec}.\\

A NC probe with frequency in this range will dress up a naked singularity of mass $-|M|$ as a black hole with the parameters
\begin{equation}
\label{dressed}
 M'= |M| \left[ \frac{4 L_{NC}|M| }{l^2 \omega} - 1 \right ] \qquad J'=0,  \qquad r_+ = l \sqrt{M'}, \qquad \textrm{and} \qquad r_- = 0,
\end{equation}
whose metric would be given by \eqref{eqbtzmetric}. Note that the above analysis can be extended for $m=0$ as well, although in that case, the expression for $g$ in \eqref{eqng} has to be analyzed with care. The above analysis provides a domain in the parameter space within which the noncommutativity dresses the naked singularity by acquiring a new horizon at $r_+=l\sqrt{M^\prime}$.

\section{Properties of the
dressed singularity}

%%%%%%%%%%%%%%%%%%%%%%%
\subsection{Thermodynamical properties and entropy}
%%%%%%%%%%%%%%%%%%%%%%%

In the previous section we have found a frequency range of the massive NC probe for which  the dressed singularity has parameters described by \eqref{dressed}. In this section we shall discuss the physical properties of the dressed singularity.\\

Let us first note that our analysis is valid only in the semi-classical limit, where the mass $|M|$ is typically much greater that the Planck mass and any other energy scale in the problem. This ensures that the quantum back-reaction effects can be ignored. Under this assumption, and for the NC probes satisfying the condition \eqref{ec}, we find that the mass of the dressed singularity $M'$ would also be much greater than the Planck mass. Thus the dressed singularity would be described by a massive spinless BTZ black hole with parameters given by \eqref{dressed} and it would be amenable to semi-classical analysis.\\

In view of these observations, the dressed singularity would satisfy all the laws of black hole thermodynamics \cite{Kim, Singh}. In particular, it would have a Bekenstein-Hawking entropy given by 
\begin{equation}
S=\frac{A}{4}=\frac{\pi r_+}{2}.
\end{equation}
Using \eqref{dressed} we obtain the entropy as
\begin{equation}
S=\frac{\pi}{2}l\sqrt{\left|M\right|}\sqrt{\frac{4 L_{NC}|M| }{l^2 \omega} - 1}.
\end{equation}
For any non-zero $L_{NC}$, it follows from \eqref{ec} that the entropy of the dressed singularity is always positive.

\subsection{QNM modes and connection to CFT}

 So far we have shown that under the influence of a NC probe, a naked singularity is dressed up as a BTZ black hole with certain parameters. We now turn to the question as to how an asymptotic observer can estimate the parameters of the dressed singularity. We propose that the quasi-normal modes (QNM) \cite{Birmingham:2001hc,Cardoso:2001hn,Birmingham:2001pj} 
  of the dressed singularity can be used for this purpose. For the purpose of simplicity, we shall consider the scalar field QNM here, although the analysis can be easily extended to the fermionic or vector field QNM as well.\\ 

In order to proceed, we couple a commutative massless scalar field with frequency $\Omega$ to the dressed singularity and solve the corresponding Klein-Gordon equation with the QNM boundary conditions. Since the dressed singularity is a BTZ black hole with mass $M'$ and angular momentum $J'$, the QNM boundary conditions state that the field is purely ingoing at the horizon and it vanishes at infinity. With these boundary conditions, the QNM frequencies are obtained as \cite{Birmingham:2001hc}
\begin{equation}
\Omega_{L/R}=\pm\frac{m}{l}-2i\left(\frac{r_+ \mp r_-}{l^2}\right)(n+1)=\pm\frac{m}{l}-2i\frac{\sqrt{M'l^2\mp \left|J'\right|l}}{l^2}(n+1),
\end{equation}
where $m\in \mathbb{Z}$ is the angular momentum of the scalar field and $n$ is an integer and $L,R$ corresponds to the left and right moving modes respectively. An asymptotic observer can measure these frequencies and thereby obtain the geometric parameters of the dressed singularity.\\

Next we discuss the dual CFT description of the dressed singularity in the context of the QNM. Physically the QNM describe the perturbation of the metric outside the black hole, whereas in the dual description, the perturbations of the CFT are also described by the same QNM frequencies. These CFT's exist at finite temperatures given by
\begin{equation}
T_{L/R}=\frac{r_+\mp r_-}{2\pi l^2}=\frac{\sqrt{M'l^2\mp \left|J'\right|l}}{2\pi l^2}
\end{equation}
The AdS/CFT duality in this context is also supported by the Sullivan's theorem \cite{sul1,sul2,sul3} which states that for a certain class (geometrically finite) of hyperbolic manifolds, there is a one-to-one correspondence between the hyperbolic structure in the interior and the conformal structure at the boundary. It has been shown that the Euclidean continuation of the BTZ black hole satisfies all the requirements of the Sullivann's theorem. Since the dressed singularity has the geometric structure of a BTZ black hole, it admits an AdS/CFT duality at the kinematical level.\\

By taking into account eq.\eqref{rewrite}
%\footnote{We use $$M\longrightarrow M(1+Mf), \quad J\longrightarrow gM^2$$} 
we can determine the QNM and the temperature scale for our dressed situation as
\begin{equation}
\Omega_{L/R}=\pm\frac{m}{l}-2i\frac{\sqrt{M(1+Mf)l^2\mp \left|g\right|M^2l}}{l^2}(n+1)
\end{equation}
\begin{equation}
T_{L/R}=\frac{\sqrt{M(1+Mf)l^2\mp \left|g\right|M^2l}}{2\pi l^2}
\end{equation}

%%%%%%%%%%%%%%%%%%%%%%%%%%%%%%%%%%%%%%%%%%%%%%%%%%%%%%%%%%
\subsection{Naked singularity and quantum completeness}%%%
%%%%%%%%%%%%%%%%%%%%%%%%%%%%%%%%%%%%%%%%%%%%%%%%%%%%%%%%%%

Naked singularities are ``part of the spacetime'' where the curvature tensor and the energy density are ill defined due to possible accordance of infinities \cite{Waldbook}. However, in the spirit of weak cosmic censorship hypothesis, proposed by Penrose \cite{Penrose:1969pc, Penrose:1900mp, Penrose:1999vj}, one can ignore this problem if such singularities are ``hidden'' behind an event horizon, since no causal signal can reach an outside observer from the questionable region.  The aforementioned reasoning is natural in a completely classical setting, but it is already known \cite{recent, recent1} that if one takes the quantum effects into account one can indeed dress the naked singularity with an event horizon, thus enforcing the weak cosmic censorship.\\

We used NC geometry as a model for describing quantum gravity effects. In the previous subsection we saw that introducing noncommutativity dresses the naked singularity in 2+1 dimension. Namely, in 2+1 dimension the black hole solution of Einstein equation (with the cosmological constant) is given by the BTZ metric \cite{banados, heno} which is parametrized by its mass $M$ and angular momentum $J$ and if these parameters satisfy $-1<M<0$ and $J=0$ we are dealing with a continuous sequence of naked singularities (point particle sources) at the origin. This singularity originates from a topological obstruction  since the Ricci curvature in that case has a term that is proportional to the Dirac $\delta$-function in addition to the constant curvature \cite{Pantoja:2002nw}. Furthermore it can be shown that this naked singularity is not resolved via quantum scalar probe \cite{pitelli, Unver}, meaning that this space is geodesically and quantum mechanically incomplete. However we have shown that by introducing the noncommutativity one can dress the naked singularity by producing an event horizon  and one can further investigate if the effective space is quantum complete. \\

Namely, the spacetime is said to be quantum complete  if the evolution of any state is uniquely defined for all time. This leads one to examine the wave equations of test fields, extract the Hamiltonian and via von Neumann's  criteria investigate its self-adjoint properties (extensions). If the aforementioned Hamiltonian has a unique self-adjoint extension, then it will render a unique time evolution and such a system is referred to as quantum complete (or non-singular), since the test field will not ``see'' the spacetime singularity during its propagation (for more details and explicit examples the reader is referred to \cite{hofmann, reedsimon, chernof, horowitz, isibasi, isibasi1, konkowski, wald, Konkowski:2016pgi}).\\

In previous works by some of the authors \cite{kumarnormal, Juric:2018qzl} it was shown that the quantum completeness is achieved if the following condition is fulfilled 
\begin{equation}
\frac{l^2 r_{+}}{r^{2}_{+}-r^{2}_{-}}\geq 1
\end{equation}
which for $Ml\gg J$ reduces to \cite{Juric:2018qzl}
\begin{equation}
\sqrt{M}\leq l
\end{equation}
Now, taking into account eq.\eqref{aaa} and \eqref{rewrite} we rerwite $M\longrightarrow M^\prime=\left|M\right|(f\left|M\right|-1)$  and obtain
\begin{equation}
f\leq \frac{l^2}{\left|M\right|^2}+\frac{1}{\left|M\right|}
\end{equation}
which combined with \eqref{uvjet} finally gives
\begin{equation}\label{theuvjet}
\frac{1}{\left|M\right|}\leq f\leq \frac{l^2}{\left|M\right|^2}+\frac{1}{\left|M\right|}
\end{equation}
Condition \eqref{theuvjet} gives a bound on $L_{NC}$ for which the naked singularity is dressed and for which the corresponding space is quantum complete for a scalar probe. \\

%%%%%%%%%%%%%%%%%%%%%%%%%%
\section{Final remarks}%%%
%%%%%%%%%%%%%%%%%%%%%%%%%%

The weak cosmic censorship hypothesis plays a fundamental role in cosmology. In this paper we have considered a toy model of 2+1 gravity in the presence of a negative cosmological constant which admits naked singularities. There are two aspects of the Planck scale physics which play important role in our analysis. First aspect is that noncommutativity generically arises at the Planck scale \cite{dop1,dop2}. We have considered a $\kappa$-deformed type of noncommutativity as that is naturally associated with black hole physics \cite{ohl}. It is also expected that the physical processes occur at very high energies at the Planck scale. This is the second feature which is relevant for our analysis.\\

Using a by now well established notion of noncommutative duality \cite{ncbtz, ncbtz1, ncbtz2, ncbtz3}, we have shown that a naked singularity in our toy model will generically be hidden under event horizon(s), provided that the probe frequency $\omega$ does not belong to the range $0 < \omega < |m|/l$. In particular, if the probe has very high energy as given in \eqref{ec}, then the resulting dressed object is a massive spinless BTZ black hole with a single event horizon. For intermediate energies the naked singularity is dressed as a massive spinning BTZ black hole. Only for very low energy probes, the singularity is not dressed. Such probes are however highly improbable at the Planck scale where physical processes typically occur at high energies. Thus for typical noncommutative probes, our analysis is consistent with the weak cosmic censorship hypothesis.\\

The noncommutative duality clearly shows that the dressed singularity is geometrically equivalent to a BTZ black hole. We can therefore immediately obtain the physical properties of the dressed singularity. In particular, such a dressed singularity would obey all the laws of black hole thermodynamics. In addition, when perturbed by a suitable field, it would give rise to the usual quasi-normal modes of a BTZ black hole with the usual holographic interpretation. Finally, we show that the spacetime of the dressed singularity can even be quantum mechanically complete depending on the nature of the probe.\\

In addition to the dressed singularity, our analysis results in a very light scalar as a byproduct of the duality transformation. Such a light scalar is of great significance in cosmology, especially in the context of quintessence \cite{quint}. However, the details of the cosmological implications of the light scalar resulting from our analysis is beyond the scope of this work.\\

We end this paper with few remarks. First, it may be noted that a large class of string theories contain a BTZ factor in the near-horizon geometry \cite{maldastrom,physrept,peet,cvetic,skenderis}. Thus, it is plausible that our analysis holds in a much broader context as described by these string theories. Secondly, in this paper we have not directly considered the quantum aspects of the noncommutative probe, in particular we have not considered backreaction effects of the test particles. It would be interesting to see what kind of physical effects  would such quantum backreactions invoke, and we plan to investigate this more in the future. Finally, we note that the analysis carried out in this paper might be extended to 3+1 dimensional spacetime
to investigate whether noncommutativity can dress a naked singularity in this more realistic situation  and in case of a positive result, to study  properties of the object obtained in this way \cite{Ciric:2017rnf,Ciric:2019uab, novi}. \\

\noindent{\bf Acknowledgment}\\
 This work was partially supported by the H2020 CSA Twinning project No.~692194, RBI-T-WINNING and by the European Union through the European Regional Development Fund -- the Competitiveness and Cohesion Operational Programme (KK.01.1.1.06).\\

%%%%%%%%%%%%%%%%%%%%%%%%%%%%%%%
%%%%%%%%%%%%%%%%%%%%%%%%%%%%%%%
%%%%%%%%%%%%%%%%%%%%%%%%%%%%%%%

\end{document}